\documentclass[showpacs,preprintnumbers,amsmath,amssymb,twocolumn,showkeys]{revtex4}
\usepackage{graphicx}   
\usepackage{dcolumn}    
\usepackage{bm}         
\usepackage{ifpdf}
\usepackage{graphicx,amssymb,lineno}
\usepackage{amssymb,lineno,amsfonts}
\usepackage{graphicx}   
\usepackage{dcolumn}    
\usepackage{bm}         
\usepackage{bbm}
\usepackage{mathrsfs}
\usepackage{upgreek}
\usepackage{mathtools}
\begin{document}

\title{Storing entanglement of nuclear spins via Uhrig Dynamical Decoupling}

\author{Soumya Singha Roy$^1$, T. S. Mahesh$^1$ and G. S. Agarwal$^2$}
\email{ss.roy@iiserpune.ac.in, mahesh.ts@iiserpune.ac.in, girish.agarwal@okstate.edu}
\affiliation{$^1$Indian Institute of Science Education and Research, Pune 411008, India \\
$^2$Department of Physics, Oklahoma State University, Stillwater, OK 74078, USA}

\date{\today}

\begin{abstract}
{
Stroboscopic spin flips have already been shown to prolong the coherence times of
quantum systems under noisy environments.  Uhrig's dynamical decoupling
scheme provides an optimal sequence for a quantum system interacting
with a dephasing bath.  Several experimental demonstrations have already
verified the efficiency of such dynamical decoupling schemes
in preserving single qubit coherences.  In this work we describe the
experimental study of Uhrig's dynamical decoupling in preserving 
two-qubit entangled states using an ensemble of spin-1/2 nuclear pairs
in solution state.
We find that the performance of odd-order Uhrig sequences
in preserving entanglement is superior to both even-order 
Uhrig sequences and
periodic spin-flip sequences.  We also find that there exists an optimal order of 
the Uhrig sequence using which a singlet state can be stored at high correlation
for about 30 seconds.  
}
\end{abstract}

\maketitle

\section{Introduction}
Harnessing the quantum properties of physical systems has 
several potential applications,
particularly in information processing, secure data communications, 
and quantum simulators \cite{chuangbook}.
It is believed that such quantum devices may play an important role in
future technology \cite{milburn}.  But their
physical realization is
challenging mainly because of decoherence - the decay of
the coherent states due to interaction with the
surrounding environment \cite{joos,schlosshauer}.
Therefore it is important to minimize the effects of decoherence
using suitable perturbation on the quantum system \cite{viola99}.
One technique known as `dynamical decoupling' involves 
protecting the quantum state from decoherence by driving 
the system in a systematic manner 
such that the effective interactions with the environment at different 
instants of time cancel one another.

Preserving nuclear spin coherences
by spin flips at regular intervals was long been known in nuclear magnetic
resonance (NMR) as the famous Carr-Purcell-Meiboom-Gill (CPMG) sequence
\cite{carr,meiboom}.  
The CPMG sequence is widely used in NMR to 
measure the transverse relaxation time constants in the presence of 
spatial inhomogeneity of the static magnetic field and temporal fluctuations 
in the local fields arising due to molecular motion \cite{LevBook}. 
The sequence involves a set of N $\pi$ pulses uniformly distributed
in a duration $[0,T]$ at time instants $\{t_1,t_2,\cdots,t_N\}$.
Assuming instantaneous $\pi$ pulses, $j^\mathrm{th}$ time instant
is linear in $j$
\begin{eqnarray}
t_j^\mathrm{CPMG} = T \left( \frac{2j-1}{2N} \right).  
\end{eqnarray}
Of course, in practice the $\pi$ pulses do have finite duration 
owing to the limited power of electromagnetic irradiation 
generated by a given hardware. 
Further, the constant time period between these spin flips should ideally
be shorter than the correlation time of the spin-bath interaction.
Even this delay is limited by the maximum duty-cycle that is
allowed for the hardware.  Dynamical decoupling with such bounded
controls have also been suggested \cite{viola03,gordon,jacob10,pasini10}.
For instance Hao et al. have been able to calculate, 
using a particular type of atomic systems, 
the maximum delay between spin-flips in order
to efficiently suppress decoherence due to a bath with a finite 
cut-off frequency \cite{liang}.
By studying the efficiency of the decoupling as a function of 
the CPMG period often it is possible to extract valuable informations 
about molecular dynamics and such studies are broadly categorized under
`CPMG dispersion' experiments \cite{palmer}.

Uhrig generalized the CPMG sequence by considering an
optimal distribution $\{t_1,t_2,\cdots,t_N\}$ of $N$ spin flips
in a given duration $[0,T]$ of time
that provides most efficient dynamical decoupling \cite{uhrig}.
Using a simple dephasing model, Uhrig proved that the 
time instants should vary as a squared sine bell:
\begin{eqnarray}
t_j = T \sin^2\left(\frac{\pi j}{2N+2} \right).
\label{tj}
\end{eqnarray}

UDD works well in systems having a high-frequency dominated
bath with a sharp cutoff \cite{Pasini,Biercuk,BiercukPRA}.
On the other hand when the spectral density of the bath has
a soft cutoff (such as a broad Gaussian or Lorentzian), the CPMG sequence was found to 
outperform the UDD sequence 
\cite{duuhrig,dieteruhrig1,Cywinsky,Lange,Barthel,Ryan,sagi}. 
Suter and co-workers have studied these different regimes and
arrived at optimal conditions for the dynamical
decoupling \cite{dieteruhrig2}. 

Recently Agarwal has shown using theoretical and
numerical calculations that even 
entangled states of two-spin systems can be stored more efficiently
using UDD \cite{agarwal}.  
Since entangled states play a central role in QIP, 
teleportation, data encryption, and so on,
saving entanglement is crucial in physical
realization of quantum devices
\cite{chuangbook}.  More recently dynamical decoupling on an 
electron-nuclear spin-pair in a solid state system 
has been shown to prolong the pseudoentanglement lifetime by
two orders of magnitude \cite{duprl}.

While much of the experimental effort has been on testing the loss of coherence due to T$_2$ processes, 
ours is one of the first experiments where we study not only the loss of coherences, but also the
loss of entanglement due to both T$_1$ and T$_2$ processes. 
Though newer sequences have been suggested to decouple both of these
processes, these are yet to be studied experimentally \cite{uhrigprl09,lidarprl10}.
Since we have developed experimental techniques where we can prepare Bell states with high fidelity 
and characterize these states with high precision \cite{maheshjmr10,maheshpra10}, 
we explore the utility of different dynamical decoupling sequences on
systems wherein both T$_1$ and T$_2$ relaxations are significant.
In the following we describe preparation of Bell states
using a two-qubit NMR system and then describe 
the comparative 
performances of CPMG and UDD on such entangled states.

\section{Preparation and Storage of Entanglement}
\begin{figure}
\includegraphics[trim = 20mm 40mm 20mm 40mm, clip, width=6.5cm]{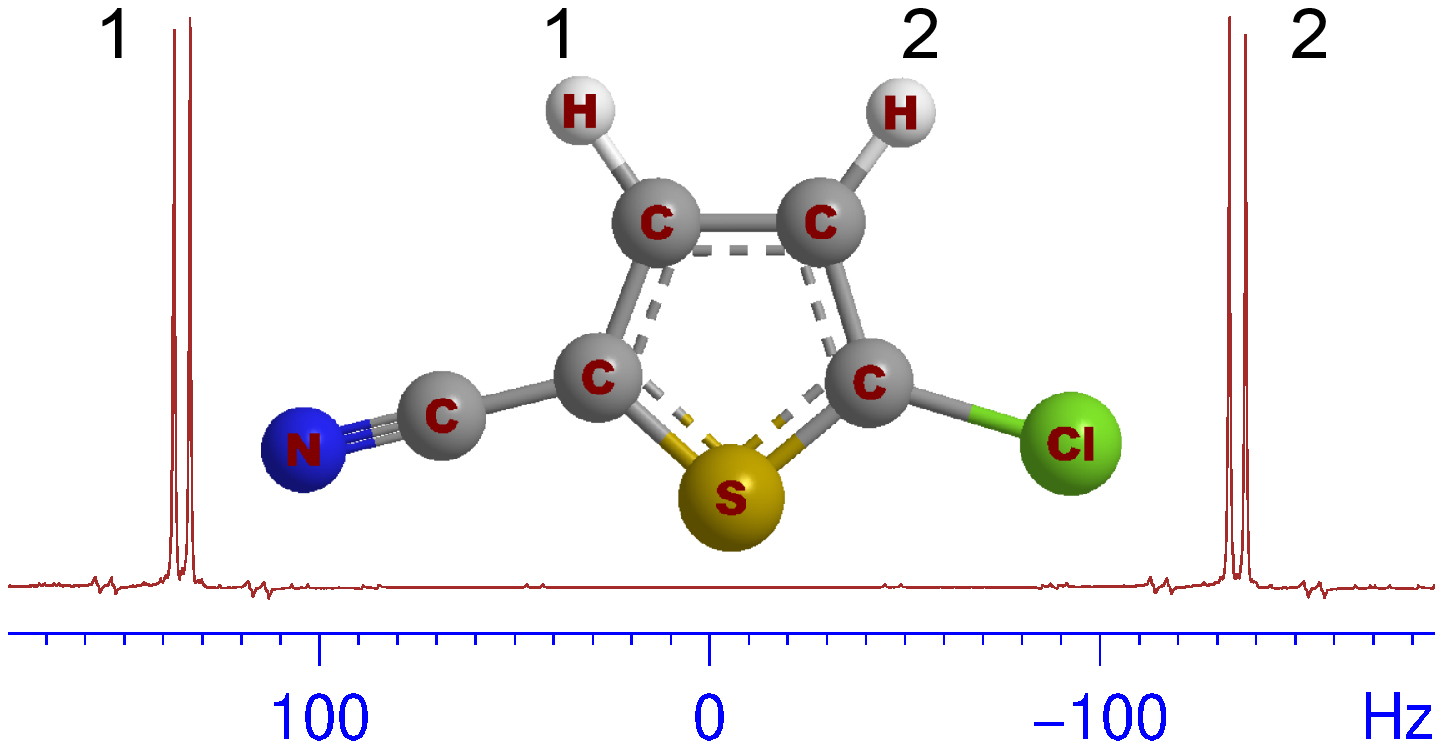} 
\caption{(Color online) The $^1$H NMR spectrum and the molecular structure
of 5-chlorothiophene-2-carbonitrile.
}
\label{fig1} 
\end{figure}

We study storage of entanglement by dynamic decoupling
on a pair of spin-1/2 nuclei using liquid state NMR techniques.
The sample consisted of 5 mg of 5-chlorothiophene-2-carbonitrile 
dissolved in 0.75 ml of dimethyl sulphoxide (see Figure \ref{fig1}).
The two protons of
the solute molecule differ in the Larmor frequency by 
$\Delta \nu = 270.4$ Hz and have an indirect spin-spin coupling 
constant of $J = 4.1$ Hz.
The $T_2$ relaxation time constants for the two protons are about 2.3 s and
the $T_1$ relaxation time constants are about 6.3 s.
All the experiments are carried out in a Bruker 500 MHz NMR 
spectrometer at an ambient temperature of 300 K.

\subsection*{Pseudopure States}
An ensemble of nuclear spin-systems in thermal equilibrium
at room temperature $T$ exists in a  mixed state at
high spin temperature, since the Zeeman energy gaps $\Delta E$ are 
much smaller than the Boltzman energy $kT$.  Therefore the
general state of an ensemble of spin-1/2 nuclear pairs can be described
by the density matrix
\begin{eqnarray}
\rho_\mathrm{eq} = \frac{1}{4}{\mathbbm 1} + \epsilon \rho_\Delta,
\label{rhoeq}
\end{eqnarray}
where the identity part is usually ignored and the traceless 
matrix $\rho_\Delta$ is called the deviation density matrix.
The dimensionless quantity  $\epsilon \sim \Delta E / kT$ and
have magnitudes $\sim 10^{-4}$ for protons in 
currently available magnetic fields
at ordinary room temperatures.
Using a combination of unitary and non-unitary processes,
it is possible to transform the above state to 
a pseudopure $\vert \psi \rangle$ state given by
\begin{eqnarray}
\rho_\mathrm{pps} = \frac{1}{4}(1-\epsilon){\mathbbm 1} + 
\epsilon \vert \psi \rangle \langle \psi \vert
\label{rhopps}
\end{eqnarray}
\cite{corypps}.
Here $\epsilon$ is a measure of the magnetization retained in
the pseudopure state. All the kets in the rest of the article must 
be understood as pseudopure states.

\subsection*{Preparation of pseudoentangled States}
High fidelity entangled states are prepared via long
lived singlet states in a procedure described by Soumya et al
\cite{maheshpra10,maheshjmr10}.  The method is briefly explained in the
following for completeness.

The Hamiltonian for an ensemble of spin-1/2 nuclear pairs of same isotope, in 
the RF interaction frame, can be expressed as
\begin{eqnarray}
{\cal H^{\mathrm{eff}}} =  h \left[ 
         \frac{\Delta\nu}{2}  I_z^1 
         - \frac{\Delta\nu}{2}I_z^2 
         +  J I^1 \cdot I^2
         +  \nu_{12} I_x^{1,2}
          \right].
\label{heff}         
\end{eqnarray}
Here $h\Delta \nu$ corresponds to the difference in the Zeeman 
energy gaps of the two spins (due to the chemical shift difference), 
$I_z^k$ corresponds to the 
spin angular momentum of $k^\mathrm{th}$ spin, $J$ is the 
indirect spin-spin coupling, and $\nu_{12}$ is the amplitude
of circularly polarized component of the radio frequency field,
which is at resonance with the mean frequency of the two spins.

In the limiting case of $\Delta \nu \rightarrow 0$, the system is
said to have magnetic equivalence, and the singlet state 
$\vert S_0 \rangle = (\vert 0 1 \rangle - \vert 1 0 \rangle)/\sqrt{2}$, and
the triplet states $\vert T_1 \rangle = \vert 0 0 \rangle$, 
$\vert T_0 \rangle = (\vert 0 1 \rangle + \vert 1 0 \rangle)/\sqrt{2}$,
and $\vert T_{-1} \rangle = \vert 1 1 \rangle$ 
form an orthonormal eigenbasis of the resulting
internal Hamiltonian \cite{LevBook} 
\begin{eqnarray}
{\cal H}_{\mathrm{eq}}^{\mathrm{eff}} = h J I^1 \cdot I^2.
\end{eqnarray}  
The equivalence Hamiltonian may be realized
by suppressing the chemical shift $\Delta \nu$ 
using a radio frequency `spin-lock' \cite{LevittJACS04}.  

\begin{figure}
\includegraphics[width=8.7cm]{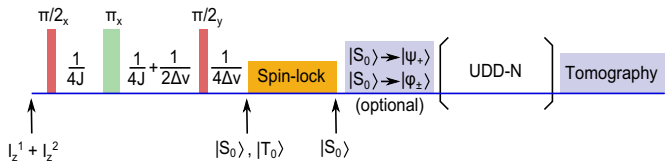} 
\caption{(Color online) NMR pulse sequence to study dynamical decoupling on
Bell states.  An incoherent mixture of singlet and triplet states
is prepared which under spin-lock purifies to singlet state.
The resulting singlet state can be converted to other Bell states.
Then dynamical decoupling sequence can be applied and the
performance of the sequence can be studied by characterizing the
residual state using density matrix tomography.}
\label{fullpulseq} 
\end{figure}

The long-lived nature of singlet states under the equivalence
Hamiltonian can be used to  prepare high-fidelity Bell states.
The experiment involves preparing an incoherent mixture of singlet
and triplet states
\begin{eqnarray}
\rho(0) = - {\bf I}^1 \cdot {\bf I}^2 \equiv
\vert S_0 \rangle \langle S_0 \vert -\vert T_0 \rangle \langle T_0 \vert
\end{eqnarray}
from the equilibrium state $I_z^1 + I_z^2$ 
by using the pulse sequence shown in Figure \ref{fullpulseq}
\cite{LevJMR06}.
After preparing
$\rho(0)$, the Hamiltonian ${\cal H}_{\mathrm{eq}}^{\mathrm{eff}}$
is imposed using RF spin-lock.
During the spin-lock $\vert T_0 \rangle$ state rapidly equilibrates
with the other triplet states. 
On the other hand, the decay constant 
of singlet state $\vert S_0 \rangle$ during the spin-lock
is much longer than the spin-lattice relaxation time constant ($T_1$)
(and hence the singlet state is known as a long-lived state) 
\cite{LevPRL04,LevittJACS04}.  The long-life is attributed to the
fact that the anti-symmetric singlet state can not be converted
to symmetric triplet states via symmetry conserving relaxation mechanisms
such as the intra-molecular dipolar interactions
\cite{LevittJCP05,LevJCP09,BodenJMR06}.  

As a result
of the long life-times under 
${\cal H}_{\mathrm{eq}}^{\mathrm{eff}}$, 
the attenuated correlation of the singlet state
\begin{eqnarray}
\langle {\bf S_0} \rangle = 
\frac{\mathrm{trace}\left[ \vert S_0 \rangle \langle S_0 \vert \cdot \rho(t) \right]}
{\sqrt{
\mathrm{trace} \left[\rho^2(t) \right]}}
\label{corr}
\end{eqnarray}
with $\rho(t)$
improves over time $t$, reaches a maximum value and ultimately
decays due to the symmetry breaking interactions
\cite{maheshjmr10}.
The spin-lock imposing the Hamiltonian ${{\cal H}_{\mathrm{eq}}^{\mathrm{eff}}}$ 
can be turned off when the maximum correlation $\langle {\bf S_0} \rangle$
is reached, and the high fidelity 
singlet state can be used for further studies.

Other Bell states can be obtained easily from the singlet
state:
\begin{eqnarray}
\vert S_0 \rangle & \xrightarrow{\mathrm{e}^{i\pi I_z^1}}    &  \vert \psi_+ \rangle = (\vert 01 \rangle + \vert 10 \rangle)/\sqrt{2}, \nonumber \\
\vert S_0 \rangle & \xrightarrow{\mathrm{e}^{i\pi I_x^1}}               & \vert \phi_- \rangle = (\vert 00 \rangle - \vert 11 \rangle)/\sqrt{2}, \nonumber \\
\vert S_0 \rangle & \xrightarrow{\mathrm{e}^{i\pi I_x^1} \cdot \mathrm{e}^{i\pi I_z^1}} & \vert \phi_+ \rangle = (\vert 00 \rangle + \vert 11 \rangle)/\sqrt{2}.
\end{eqnarray}
The $z$-rotation in the above propagators can be implemented by 
using chemical shift evolution for a period of $1/(2\Delta \nu)$, and
qubit selective $x$-rotation can be
implemented by using radio frequency pulses
\cite{maheshpra10}.

Details of dynamical decoupling on the Bell states
will be described in the next sub-section.  In order to
investigate the decoupling performance one needs to 
quantify the decay of Bell states with decoupling duration.
The Bell states by themselves are inaccessible to
macroscopic observables, but
can indirectly be detected 
transforming to observable single quantum coherences
\cite{LevPRL04,LevittJACS04}.
Alternatively, a more detailed and quantitative analysis of Bell states 
may be carried out  using density matrix tomography
\cite{maheshjmr10}. 
After measuring the density matrix, attenuated correlation can 
be evaluated for any Bell state using expressions similar to
(\ref{corr}).

In the following we describe experimental implementations of
dynamical decoupling after preparing such entangled states.

\begin{figure}
\includegraphics[width=8.5cm]{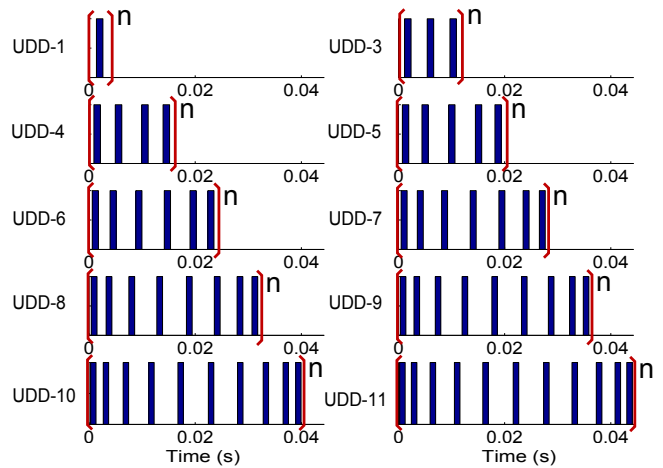} 
\caption{(Color online) Pulse sequences for various orders of Uhrig Dynamical
Decoupling.  Note that both UDD-1 and UDD-2 are equivalent to CPMG.
The time instants are calculated according to the expression
(\ref{tj}), with $N$ being the order of UDD and the
total period $T = N \times 4.0272$ ms.
}
\label{fig2} 
\end{figure}

\subsection*{Dynamical Decoupling}
As described earlier, the UDD scheme consists of a sequence of
spin flips placed at time instants given by the expression
(\ref{tj}).
Instead of applying the Uhrig's formula for the
entire duration of decoupling, we apply the formula for a short
time interval ($T$) consisting of a small number ($N$) of pulses
and then repeat the sequence. 
Figure \ref{fig2} shows pulse sequences for various orders
of Uhrig Dynamical Decoupling (we refer to an N-pulse 
UDD sequence as UDD-N).
Note that UDD-1 (and UDD-2) are equivalent to CPMG sequence,
in which repeating segment consists of $[\tau_\mathrm{CPMG} - \pi - \tau_\mathrm{CPMG}]$.
In our experiments,  $\tau_\mathrm{CPMG}$ was set to 2 ms and
the duration $\tau_{\pi}$ of the $\pi$ pulse was 27.2 $\mu$s.
The total duration of UDD-N was set to $T = N(2\tau_\mathrm{CPMG}+\tau_{\pi})$,
such that for an extended period of time, the total number of 
$\pi$ pulses remains same irrespective of the order of UDD.  Only
the distribution of $\pi$ pulses varies according to the order 
of UDD.  For example in one second of decoupling, there will be about 
250 $\pi$ pulses in all UDD-N.  Our investigation thus helps in studying
the efficiency of decoupling over a fixed duration of time
for a given number of $\pi$ pulses 
dispersed according to different orders of UDD.

\begin{figure}
\hspace*{-1cm}
\includegraphics[width=9.5cm]{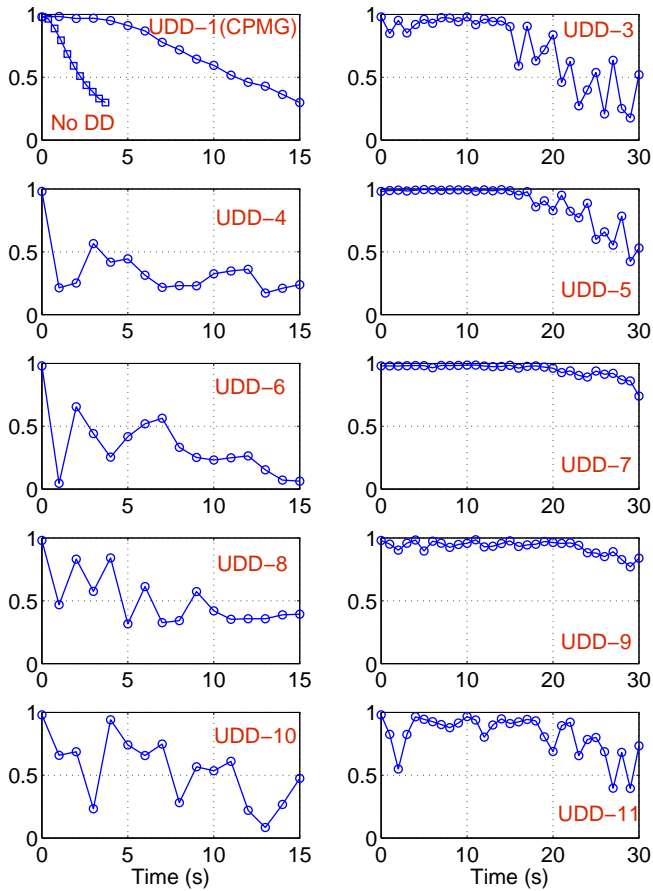} 
\caption{(Color online) Experimental correlations (circles) of singlet state as a function of 
decoupling duration of various orders of UDD.  Also shown in the
top-left figure is the correlation decay under no dynamical decoupling (squares).
}
\label{uhrigall} 
\end{figure}

Now we describe the performances of UDD-N on 
the singlet state which was prepared as explained before
(see Figure \ref{fullpulseq}).
After applying UDD-N for a fixed duration of time, we carried out
density matrix tomography and evaluated the correlation of the
preserved state with theoretical singlet density matrix.  The correlations
for various orders of UDD are displayed in Figure \ref{uhrigall}.
As can be seen from the figure, the singlet state can be
preserved for longer durations by UDD-1 (CPMG) than no-decoupling.
It is also clear that all even-order UDD sequences result in significant
fluctuations in the correlation of the singlet state.
However, the odd order UDD preserve the singlet state for tens of seconds.  
For example, the correlation of the singlet state under UDD-7 at all the
sampled time points till 20 seconds is above 0.96.
This rather surprising even-odd behavior is likely due to the differences
in the performances of the even and odd ordered sequences against
the spatial inhmongeneity of the RF pulses.

\begin{center}
\begin{figure}
\includegraphics[width=4.5cm,angle=-90]{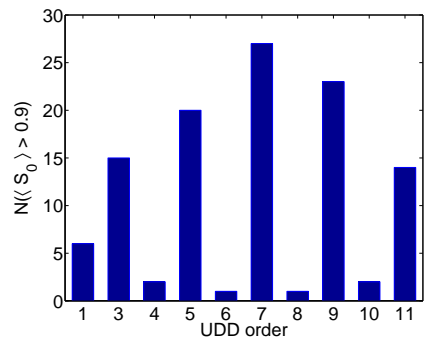} 
\caption{(Color online) The number of time instants at which the correlation 
exceeded 0.9 for various orders of UDD-N.
}
\label{bar1} 
\end{figure}
\end{center}

One way to quantify the efficiency 
of dynamical decoupling under various orders of UDD 
in figure \ref{uhrigall}, is by counting the number
of time instants in which the correlation of the preserved state
exceeds a given threshold.
The bar plot in Figure \ref{bar1} compares the number of time
instants during decoupling under various orders of UDD 
in which the correlation of the singlet state exceeded 0.9.
It can be seen that there exists an optimal order of
UDD (for a given $\tau_\mathrm{CPMG}$ and $\tau_\pi$), which performs the 
most efficient decoupling.  
The optimality may be because of the finite width of the $\pi$ pulse.
In a CPMG sequence the $\pi$ pulses are uniformly dispersed, 
while in Uhrig
sequence  the $\pi$ pulses are more crowded at the terminals 
(beginning and ending) of the sequence.
For example, if there are too many
$\pi$ pulses, Uhrig's formula will lead to an overlap of 
pulses.  Experimentally, the overcrowding of $\pi$ pulses 
may also lead to RF heating of the sample and the probe.
Thus the performance of the UDD sequence does not grow indefinitely
with the order of the sequence, but instead will fall beyond
a certain order.
In our experimental setting, we find that UDD-7 is the optimal
sequence for storing the singlet state.
There are recent suggestions for decoupling using finite pulses, 
however these are yet to be studied experimentally
\cite{uhrignjp10,uhrigcomm}.

\begin{figure}
\includegraphics[width=6.5cm]{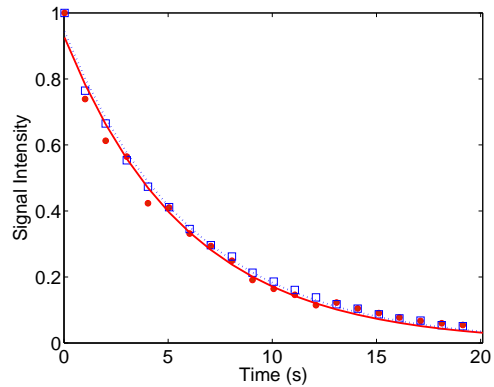} 
\caption{(Color online) The decay of the singlet spin-order
measured by converting it into observable
single quantum magnetizations.  The decay was
studied under CPMG sequence (squares) as well as under
Uhrig sequence (filled circles).
The dashed and the solid line correspond to
the exponential fits for CPMG and UDD-7 data points
respectively.
}
\label{fig3} 
\end{figure}

It can be noticed that the attenuated correlation
(expression (\ref{corr})) is insensitive
to the decay of the overall magnetization ($\epsilon$
in (\ref{rhopps})), 
but simply measures the overlap between $\rho_\Delta$
and the theoretical density matrix $\vert \psi \rangle \langle \psi \vert$.  
An alternate method is to monitor the decay of magnetization
(i.e., $\epsilon$) under dynamical decoupling.  As already
mentioned, singlet state itself can not be measured directly,
but can be converted to observable magnetization by using
a chemical shift evolution for a duration $1/(4\Delta \nu)$ 
followed by a $\left( \frac{\pi}{2} \right)_{x(y)}$ pulse. 
Intensity of the resulting signal  as a function of the 
duration of dynamical decoupling is shown in Figure \ref{fig3}.
As can be seen, UDD-7 is no better than CPMG in preserving the
overall spin-order.  In fact the decay constant for CPMG 
and UDD-7 are 6.1 s and 5.9 s respectively.

\begin{center}
\begin{figure}
\hspace*{-.4cm}
\includegraphics[width=9.1cm]{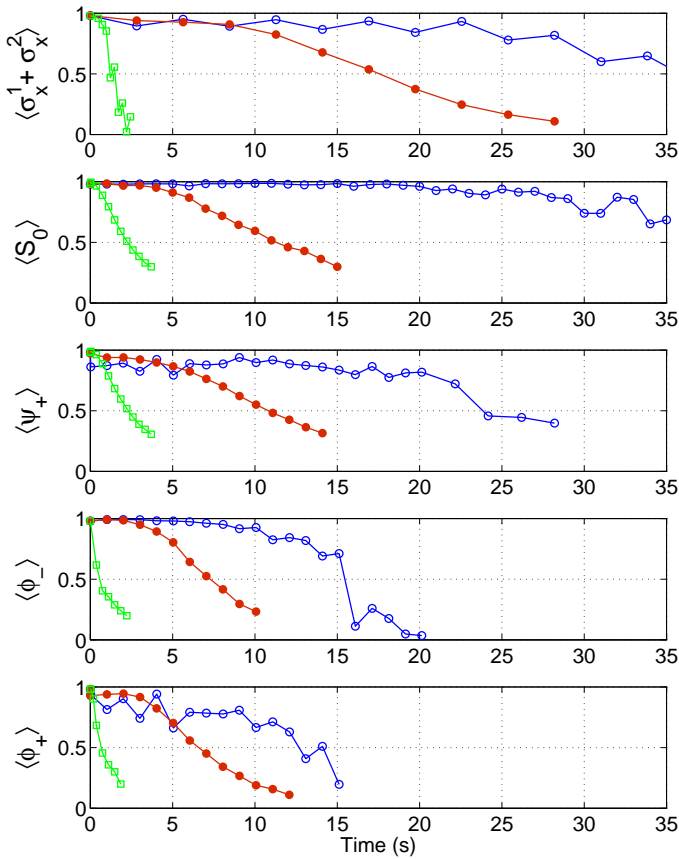} 
\caption{(Color online) Experimental correlations of the product state
and various Bell states as a function of duration under
(i) no decoupling (open squares), (ii) CPMG sequence (filled
circles), and (iii) UDD-7 (open circles).
}
\label{fig4} 
\end{figure}
\end{center}

Now we compare the efficiency of the optimal sequence UDD-7 
with UDD-1 (CPMG) for preserving product state ($\sigma_x^1 + \sigma_x^2$)
and other Bell states.
Figure \ref{fig4} shows the variation of correlation of 
product states and the Bell states as a function of the 
decoupling duration.  Here, after preparing each of the initial state,
the dynamical decoupling was applied for a fixed duration of
time.  To monitor the correlation, we have carried out the
density matrix tomography as described earlier \cite{maheshjmr10}.
In the case of no decoupling,
we observe a rapid decay of the correlation.  The
UDD-1 (CPMG) sequence shows some improvement in the storage time.  
However, UDD-7 clearly exhibits much longer storage times than 
the CPMG sequence.
The superior performance of UDD-7 on the singlet
state compared to other Bell states is
presumably because of its antisymmetric property described
in section II.

\section{Conclusions}
We have studied the efficiencies of CPMG as well as Uhrig
dynamical decoupling sequences on 2-qubit Bell states both in terms of
magnetization as well as in terms of correlation decay.  While
the Uhrig sequence is no better than CPMG sequence in terms of preserving
the overall magnetization (or spin order), it clearly outperforms the CPMG sequence in
preserving the correlation of the entangled as well as non-entangled 
states.  We summarize three important features: 
(i) the even-order UDD sequences result is fluctuations in correlations,
(ii) the odd-order UDD sequences out-perform the CPMG sequence, and
(iii) there exists an optimal length for the odd order
UDD sequence which exhibits most efficient decoupling.
In our case,
UDD-7 of 28.2 ms duration appeared to outperform all other sequences
of both lower and higher orders.  
We are carrying out
investigations into the effects of other experimental issues like
RF inhomogeneity, resonance off-set, errors in calibration of
pulse angle etc.
These considerations may help in the theoretical and practical 
understanding of the optimal decoupling schemes.

\acknowledgments
GSA thanks Director, IISER-Pune for the invitation to work at the 
Institute which led to this collaboration.
TSM acknowledges useful discussions with Prof. Dieter Suter,
Prof. Anil Kumar, and Dr. Karthik
Gopalakrishnan.
The use of 500 MHz NMR spectrometer at NMR Research Center, IISER-Pune
is acknowledged.

\end{document}